\input epsf 
\documentstyle[twocolumn,psfig]{article}
\title{SC(R)$^3$:  Towards Usability of Formal Methods}

\author{Marsha Chechik     \\
        \mbox{} \\
        University of Toronto\\
	Department of Computer Science\\
        \vspace{2cm}
        \mbox{}        }

\date{}

\begin{document}

\newcommand{\cascon}{{CASCON}}
\newcommand{\scrrr}{SC(R)$^3$}
\newcommand{\horline}{\rule{6.5in}{0.02in}}
\bibliographystyle{plain}

\maketitle
\thispagestyle{empty}

\begin{abstract}

This paper gives an overview of \scrrr\ -- a toolset designed to
increase the usability of formal methods for software development.
Formal requirements are specified in \scrrr\ in an easy to use and review 
format, and then used 
in checking requirements for correctness and in verifying consistency between annotated
code and requirements.

In this paper we discuss motivations behind this work, describe 
several tools which are part of \scrrr, and illustrate their
operation on an example of a Cruise Control system.
\end{abstract}

\section{Introduction}

Researchers have long been advocating using mathematically-precise
(``formal'') specifications in software projects~\cite{parnas93a}.
These specifications aid in removing errors early in the software
lifecycle and can be reused in a variety of ways: as a reference point
during system design and during development of test cases, as
documentation during maintenance, etc.  However, there is a strong
resistance against adopting formal methods in practice, especially
outside the domain of safety-critical systems. The primary reason for
this resistance is the perception of software developers regarding the
applicability of formal methods -- these methods are considered to be
hard (require a level of mathematical sophistication beyond that
possessed by many software developers), expensive, and not relevant
for ``real'' software systems~\cite{hall90}.  Although case-studies
(e.g. \cite{clarke96a}) have shown applicability and effectiveness of
formal methods for various industrial-size systems, this perception
still remains.
Currently, most research in formal methods concentrates on improving
modeling languages and tool support to be able to specify and verify
larger and more complex problems (e.g. \cite{holzmann97,kurshan96}).
However, to facilitate a wide-spread use of formal methods, another,
complimentary approach is necessary: to improve {\em usability} of the
methods and the tools and to demonstrate the cost-effectiveness of
applying them to software systems.  We believe that the way to make
formal methods more usable is by
\vspace{-0.1in}

\begin{itemize}
\item amortizing the cost of creating formal documentation throughout the
software lifecycle, i.e. using this documentation for checking
correctness of programs, generating test cases and test environments, etc.;

\vspace{-0.1in}

\item using (or developing) easy to read and review notations (e.g.,
state-machines and tables);

\vspace{-0.1in}

\item decreasing analysis cost through automation; and

\vspace{-0.1in}

\item adopting existing technologies wherever possible.
\end{itemize}

\vspace{-0.1in}
\noindent
We are interested in specifying and verifying event-driven systems,
and chose SCR (Software Cost Reduction) to be the requirements notation
used in our project.  
 
The SCR requirements notation was developed by a research group at the
Naval Research Laboratory as part of the Software Cost Reduction
project~\cite{heninger78,heninger80}.  This notation specifies
event-driven systems as communicating state machines which move
between states as the environment changes.  The functional part of the
system requirements describes the values of the system's output
variables as a function of the system's input (event) and internal
state.  These requirements can be formally specified by providing
structured decision tables.  For each output variable, there is a
table which specifies how to compute the variable's value based on its
previous value and input events.  

Representing logical formulae using tables is a powerful way
to visualize information which gained acceptance among many
practitioners.  Table structure makes specifications easy to
write and review and allows for high-yield mechanical analysis.
Tools were developed to check mode tables of SCR for correctness with
respect to global properties using model-checking~\cite{atlee96a} and
theorem-proving~\cite{owre97}.  A group at the Naval Research Lab
developed an industrial-quality tool called {\tt SCR*} which allows
specifying and reasoning about complex systems using
SCR~\cite{heitmeyer96b}.  {\tt SCR*} performs checks to ensure
that the tables are complete and consistent.  Tool-building is not
limited to the SCR community.  For example, David Parnas and his
colleagues are working on methods and tools to document programs using
tables~\cite{parnas94b,parnas95a,peters98,abraham97}, and a group at Odyssey Research
Associates are developing {\tt Tablewise} - a tool to reason about
decision tables~\cite{hoover95}.  However, none of these tools are
aimed at {\em using} tabular requirements once they have been created.


During the past several years, we have been developing a number of
tools that use SCR requirements throughout the various stages of
software lifecycle, and have recently integrated them into a toolset
called \scrrr\ which stands for the SCR Requirements Reuse.  
\scrrr\ allows users to specify their requirements through a visual interface,
conducts simple syntactical checks, and invokes various tools to
perform analysis of software artifacts.  We have developed 
tools to check requirements for correctness and to
verify consistency between annotated code and requirements, and will
describe these activities using the Cruise Control system -- a case
study that we recently undertook.  


The rest of this paper is organized as follows:
Section~\ref{casestudy} describes requirements of the Cruise Control
system.  Sections~\ref{reqanalysis} and~\ref{codeanalysis} outline
techniques to analyze the consistency of the requirements and the
correspondence between the code and the requirements, respectively.
In Section~\ref{conclusion} we summarize our work and outline future
research directions.

\begin{table*}[t]
\begin{center}
\begin{tabular}{|l|l|} \hline
{\bf Component} & {\bf Description} \\ \hline \hline
monitored variables & quantities that influence the system behaviour \\ \hline
mode classes & sets of states (called modes) that partition the monitored \\
\ & environment's state space \\ \hline
controlled variables & quantities that the system regulates \\ \hline
assumptions & assumed properties of the environment \\ \hline
goals & properties that are required to hold in the system \\
\hline
\end{tabular}
\caption{Components of a requirements specification in the \scrrr\ notation.
\label{table_SCR}}
\end{center}
\end{table*}

\section{Requirements of Cruise Control System}
\label{casestudy}

A Cruise Control system specified by Jim Kirby~\cite{kirby88} is
responsible for keeping an automobile traveling at a certain speed.
The driver accelerates to the desired speed and then presses a button
on the steering wheel to activate the cruise control.  The cruise
control then maintains the car's speed, remaining active until one of
the following events occurs: (1) the driver presses the brake pedal;
(2) the driver presses the gas pedal; (3) the car's speed becomes
uncontrollable; (4) the engine stops running; (5) the driver turns the
ignition off; (6) the driver turns the cruise control off.  If any of
the first three events listed above occur, the driver can re-activate the
cruise control system at the previously set speed by pressing a
`resume' button.

Table \ref{table_SCR} gives an overview of the different
sections of an \scrrr\ requirements specification.  
\scrrr\ uses a simplified SCR method in which monitored and controlled variables
have just boolean values (representing predicates on inputs and
outputs), and results of intermediate computations (so called ``terms'')  are not used.
\begin{table}[htb]
\begin{center}
\begin{tabular}{|l|l|} \hline
{\bf Variable} & {\bf Description} \\ \hline \hline {\tt Ignition} &
ignition is on \\ \hline {\tt Running} & engine is running \\ \hline
{\tt Toofast} & Sp\_Vehicle $>$ MAX\_SPEED \\ \hline {\tt Brake} &
brake pedal is being pressed \\ \hline {\tt Accel} & gas pedal is
being pressed \\ \hline {\tt b\_Cruise} & cruise button is being
pressed \\ \hline {\tt b\_Resume} & resume button is being pressed \\
\hline {\tt b\_Off} & off button is being pressed \\ \hline 
{\tt speed\_slow} & Sp\_Vehicle $<$ Sp\_Desired -\\ &
THRESHOLD\footnotemark\\ \hline
\end{tabular}
\caption{Select Cruise Control monitored variables. \label{table_monitored}}
\end{center}
\end{table}
{\small
\begin{table}[htb]
\begin{center}
\begin{tabular}{|l|l|} \hline
{\bf Mode} & {\bf Description} \\ \hline \hline
{\tt Off} & vehicle's ignition is off \\ \hline
{\tt Inactive} & vehicle's ignition is on, but the\\ & cruise control is not on \\ \hline
{\tt Cruise} & the cruise control is on and can\\ & control
the vehicle's speed \\ \hline
{\tt Override} & the cruise control is on but cannot\\ & control
the vehicle's speed \\ \hline
\end{tabular}
\caption{Modes of the mode class {\tt CC}. \label{table_modes}}
\end{center}
\end{table}}
\footnotetext{The
value of THRESHOLD is not specified in \cite{kirby88}.  That document
suggests to use a ``value chosen according to how wide a speed variation
is regarded as acceptable."}
Monitored variables in our case study indicate the state of the ignition,
brake, and acceleration, the buttons operating the cruise control system,
and the speed of the car.  Table~\ref{table_monitored} shows some
monitored variables and the predicates they represent.
{\small
\begin{table*}[t]
\begin{center}
\begin{tabular}{|c||cccccccc||c|} \hline
{\bf Current} & \ & \ & \ & \ & \ & \ & \ & \ & {\bf New} \\ 
{\bf Mode} & {\tt Ignition} & {\tt Running} & {\tt Toofast} & {\tt Brake} & {\tt Accel} & 
             {\tt b\_Cruise} & {\tt b\_Resume} & {\tt b\_Off} & {\bf Mode}\\ \hline \hline
{\tt Off}      & @T & -- & -- & -- & -- & -- & -- & -- & {\tt Inactive} \\ \hline
{\tt Inactive} & @F & -- & -- & -- & -- & -- & -- & -- & {\tt Off} \\
  \      & -- &  t &  f &  f &  f & @T & -- & -- & {\tt Cruise} \\ \hline
{\tt Cruise}   & @F & -- & -- & -- & -- & -- & -- & -- & {\tt Off} \\
  \      & -- & -- & @T & -- & -- & -- & -- & -- & {\tt Inactive} \\
  \      & -- & @F & -- & -- & -- & -- & -- & -- & {\tt Inactive} \\
  \      & -- & -- & -- & @T & -- & -- & -- & -- & {\tt Override} \\
  \      & -- & -- & -- & -- & @T & -- & -- & -- & {\tt Override} \\
  \      & -- & -- & -- & -- & -- & -- & -- & @T & {\tt Override} \\ \hline
{\tt Override} & @F & -- & -- & -- & -- & -- & -- & -- & {\tt Off} \\
  \      & -- & @F & -- & -- & -- & -- & -- & -- & {\tt Inactive} \\
  \      & -- & -- &  f &  f &  f & -- & @T & -- & {\tt Cruise} \\
  \      & -- & -- &  f &  f &  f & @T & -- & -- & {\tt Cruise}\\ %
\hline
\end{tabular}

{\bf Initial Mode:}  {\tt Off} ($\sim${\tt Ignition})
\caption{Mode transition table for mode class {\tt CC} of the cruise control system.
\label{table_CC}}
\end{center}
\horline
\end{table*}}
{\small \begin{figure*}[htb]
\begin{center}
\begin{minipage}[htb]{3in}
\psfig{figure=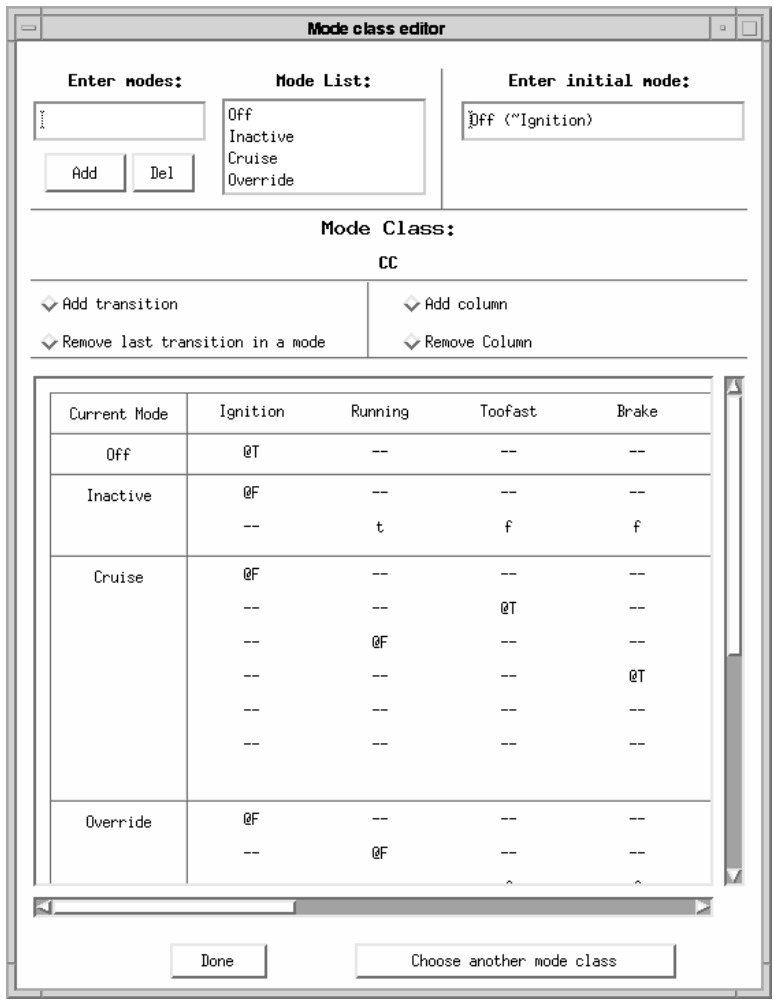,width=4in}
\end{minipage}
\caption{\scrrr\ mode class editor. \label{fig_modeclass_editor}}
\end{center}
\horline
\end{figure*}}
{\small \begin{figure*}[t]
\begin{center}
\begin{minipage}[thb]{3in}
\psfig{figure=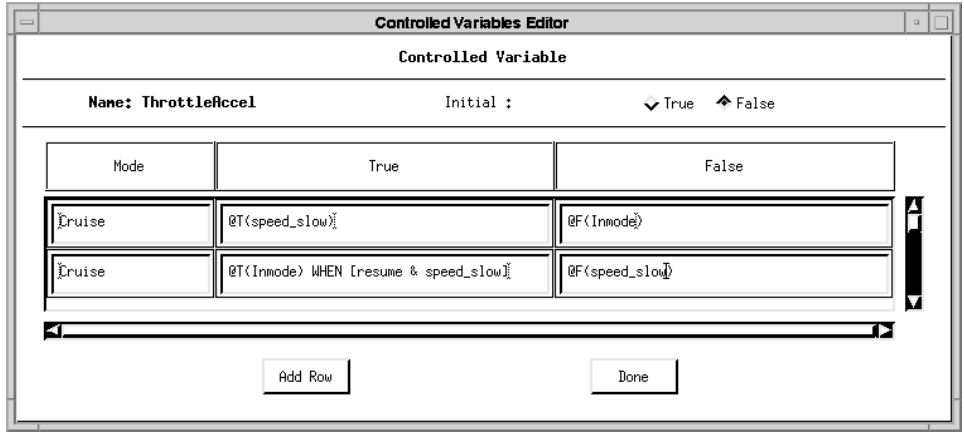,width=5in}
\end{minipage}
\caption{\scrrr\ controlled variable editor. \label{fig_control_table}}
\end{center}
\end{figure*}}
{\small \begin{table*}[htb]
\small
\begin{center}
\begin{tabular}{|l|l|l|} \hline 
{\bf Mode} & {\bf True} & {\bf False} \\ \hline \hline
{\tt Cruise} & @T({\tt speed\_slow}) & @F(Inmode) \\ \hline
{\tt Cruise} & @T(Inmode) WHEN [{\tt b\_Resume} \& {\tt speed\_slow}] & @F({\tt speed\_slow})\\ \hline
\end{tabular}

{\bf Initial:} False
\caption{Event table for the controlled variable {\tt ThrottleAccel}.
\label{table_ThrottleAccel}}
\end{center}
\horline
\end{table*}}

The Cruise Control system has one mode class {\tt CC}, whose modes are
described in Table~\ref{table_modes}.  The system is in exactly one
mode of each modeclass at all times, so we can think of modeclasses
as finite-state machines.  The mode transition table of
mode class {\tt CC} is shown in Table~\ref{table_CC}.  Each row
of the table specifies an event that activates a transition from the
mode on the left to the mode on the right.
The system starts in mode {\tt Off} if {\tt Ignition} is false,
and transitions to mode {\tt Inactive} when {\tt Ignition} {\em becomes
true}, i.e., has a value false in the current state and a value true in the
next, indicated by ``@T'' in the mode transition table.  Once in mode
{\tt Inactive}, the system remains there until {\tt Ignition} {\em becomes
false} (indicated by ``@F''), at which point it switches to mode {\tt
Off}.  The system also transitions from {\tt Inactive} to {\tt Cruise}
if {\tt b\_Cruise} {\em becomes true} while {\tt Running} is true (indicated
by ``t''), and {\tt Toofast}, {\tt Brake}, {\tt Accel}  are false
(indicated by ``f'').  Values of monitored variables indicated by
``--'' are not relevant to that particular transition, e.g., a
variable {\tt Brake} in the transition from {\tt Off} to {\tt
Inactive}.  An SCR specification defines one mode transition table for
each mode class of the system.  Such tables are entered into \scrrr\
using a mode class editor shown in Figure~\ref{fig_modeclass_editor}.

The Cruise Control system has a number of controlled variables to
control the car throttle and display messages when the car is due for
service or when it needs more oil.  For example, a variable {\tt
ThrottleAccel} is true when the throttle is in the accelerating
position and false otherwise.  The condition table for this variable
is shown in Table~\ref{table_ThrottleAccel}.  {\tt ThrottleAccel} is
only evaluated in mode {\tt Cruise}.  The first row of the table
specifies that {\tt ThrottleAccel} should become true when the speed
is too slow, and false when the system exits the mode {\tt Cruise}
(indicated as @F(Inmode)).  When the system enters {\tt Cruise}
because the user pressed the resume button, the cruise control needs
to maintain the previously set speed.  Thus, the current speed is
immediately evaluated, and if it is too slow, {\tt ThrottleAccel}
should become true, as indicated in the second row of the table.  An
SCR specification defines one table for each controlled variable of
the system.  The
\scrrr\ module for specifying these variables is shown in
Figure~\ref{fig_control_table}.

\scrrr\ allows to record assumptions of the requirements about the 
behavior of the environment.  Assumptions specify constraints on the
values of conditions, imposed by laws of nature or by other mode
classes in the system.  In particular, we can assume that the engine
is running only if the ignition is on ({\tt Running --$\gg$ Ignition}),
the car can be going ``too fast'' only if it is running
({\tt Toofast --$\gg$ Running}), and various boolean conditions
are related by enumeration.  The later category includes
predicates on the vehicle's speed ({\tt speed\_slow / speed\_ok / speed\_fast})
and buttons controlling the cruise control system ({\tt b\_Off / b\_Cruise / b\_Resume}).

The last section of requirements specification consists of the system
goals.  These are not constraints on the system behaviour but rather
{\em putative theorems } -- global properties that should hold in the
system under specification, e.g. ``the light will eventually become
green'', or ``reversing a list twice gives us the original list''.
The language for specifying these properties in \scrrr\ is an
extension of CTL (Computational Tree Logic).  CTL is a branching-time
temporal logic~\cite{clarke86} which allows quantification over some
or all possible futures.  CTL formulae are defined recursively: all
propositional formulae are in CTL; if $f$ and $g$ are in CTL, so are
${\sim}f$ (negation), $f \; \& \; g$ (conjunction), and $f \mid g$
(disjunction).  Furthermore, the universal ($A$) and the existential ($E$)
quantifiers are used alongside the ``next state'' ($X$), ``until''
($U$), ``future'' ($F$) and ``global'' ($G$) operators.  Thus, the
formula $$ AG (f \rightarrow g) $$ means that it is invariantly true
that $f$ implies $g$.  

In addition to propositional formulae, our language allows to express
properties involving SCR events, e.g.
\begin{quote}
{\bf [Property 1]} $\; \;$ If the system is in mode {\tt Override}, then it will react
to the event {\tt @F(Ignition)} by immediately going to the
mode {\tt Off}.
\end{quote}
The semantics of events @T($a$) (@F($a$)) is that $a$ is true (false)
in the current state and false (true) in the previous.  To ease
the phrasing of CTL formulas that refer to the occurrence of
conditioned events, we use unary logic connectives @T and @F to
express the SCR notions of {\em becoming true} and {\em becoming false}.

Typically, the properties we are interested in
specifying include invariants, reachability properties -- these
are important to ensure that the invariant properties are not true
vacuously, and ``progress'' (bounded liveness) properties.  For
example, some of the invariant properties of the Cruise Control system
is ``whenever the system is in mode {\tt Override} of modeclass {\tt
CC}, the system is running and the ignition is on'', formalized as $$
AG({\tt CC=Override} \rightarrow ({\tt Ignition} \; \& \; {\tt
Running}))$$
and ``predicates representing the state of the Throttle are related by
enumeration'', formalized as
$$AG({\tt ThrottleOff} \mid {\tt ThrottleMaintain}$$

\vspace{-0.3in}

$$ \mid {\tt ThrottleAccel} \mid {\tt ThrottleDecel})$$
``Progress'' properties are used to check that the system
behaves according to our expectations.  For example, one of the
``progress'' properties of the Cruise Control system is Property 1 given
above, formalized as
$$ AG({\tt CC=Cruise} \rightarrow $$

\vspace{-0.3in}

$$ AX(@F({\tt Ignition}) \rightarrow {\tt CC=Off}))$$
\newpage
Note that in this property we use a logical connective @F.

\section{Checking Consistency of Requirements}
\label{reqanalysis}

\begin{figure}[htb]
\begin{center}
\begin{minipage}[htb]{3in}
\psfig{figure=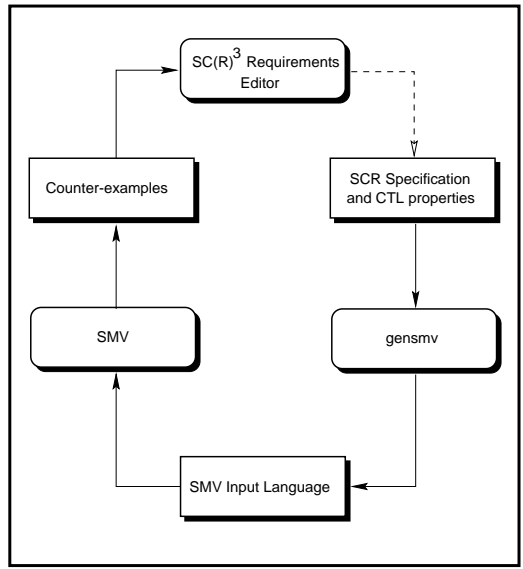,width=2.7in}
\end{minipage}
\caption{Requirements analysis with \scrrr. \label{fig_reqs}}
\end{center}
\end{figure}
System goals allow semantical checks on requirements, i.e. checks that
these goals hold in the specification of the system.  We chose to use
{\em model-checking}~\cite{clarke86} to perform these checks.  A
symbolic model-checker {\tt SMV} developed by
McMillan~\cite{mcmillan93}, uses state exploration to check if
temporal properties hold in a finite-state model.  However, we first
needed to translate the behavioral specifications of SCR into a format
accepted by {\tt SMV}, which is done using a tool {\tt gensmv}.  The
requirements analysis process is depicted in Figure~\ref{fig_reqs},
where tools and artifacts are represented by ellipses and boxes,
respectively.  In this section, we briefly describe {\tt gensmv},
outline counter-example facilities of {\tt SMV} and \scrrr, and present
results of verification of the Cruise Control system.

\subsection{Translating SCR Specifications}
{\tt gensmv}~\cite{atlee96a} was developed at the University of
Waterloo to reason about mode transition tables.  Before translating
SCR specifications, {\tt gensmv} details the mode transition tables
with information derived from environmental assumptions.  For example,
an assumption {\tt Running --$\gg$ Ignition} adds information to the
second transition from mode {\tt Inactive} to mode {\tt Cruise} (third
row of Table~\ref{table_CC}): if {\tt Running} is true, then {\tt
Ignition} should already be true.

Detailed SCR tables are translated into the {\tt SMV} input language.  In order to translate
added logic connectives @T and @F to regular CTL, accepted by {\tt SMV}, {\tt gensmv}
needs to store previous values of variables which can occur in events.  This is facilitated
by introducing additional variables in the {\tt SMV} model.  For example, in order to
reason about a property involving @F({\tt Ignition}), we need an additional variable {\tt PIgnition},
which is assigned the current value of {\tt Ignition} before the next value of this variable
is computed.  Thus, Property 1 is translated into CTL as
$$ AG({\tt CC=Cruise} \rightarrow  AX(({\tt PIgnition} \; \&$$

\vspace{-0.27in}

$$ \sim{\tt Ignition}) \rightarrow {\tt CC=Off}))$$
\noindent
We extended {\tt gensmv} to translate condition tables of controlled
variables into the {\tt SMV} modeling language~\cite{chechik98a}.
This way, we are able to reason about the entire \scrrr\ specification.

\subsection{Counter-examples}
During verification of CTL properties, {\tt SMV} explores all possible behaviors
of the model and either declares that the property holds or gives
a counter-example.  Since we wanted to make calls to {\tt SMV} completely transparent
to the user, and because of the introduction of extra variables into the
{\tt SMV} model, we found it necessary to automatically capture {\tt SMV}'s counter-examples
and translate them into the SCR format.  This translation allows the user
to easily determine where errors occur, without having to understand the
intricacies of translation between the SCR and the {\tt SMV} models.

\begin{figure}[htb]
\begin{center}
\begin{minipage}[htb]{2.5in}
\psfig{figure=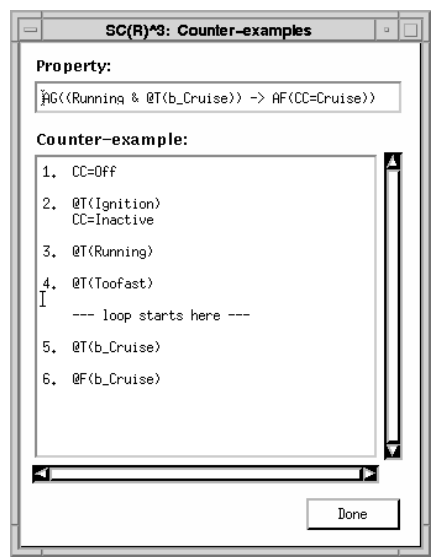,width=2.2in}
\end{minipage}
\caption{A counter-example generated during the requirements analysis. \label{fig_error}}
\end{center}
\end{figure}
\noindent
For example, if we try to check the property ``pressing the Cruise
button at any point when the system is running will turn the cruise
control system on'', formalized as $$ AG(({\tt Running} \; \& \;
@T({\tt b\_Cruise})) $$

\vspace{-0.27in}

$$ \rightarrow AF({\tt CC=Cruise})) $$
\scrrr\ reports a scenario that would violate this property, shown in Figure~\ref{fig_error}.
Thus, the system does not react to changes in {\tt b\_Cruise} if the
car is running too fast to be controlled by the automatic system.

\subsection{Checking the Cruise Control System}

{\small \begin{table*}[t]
\begin{center}
\begin{tabular}{|l|c|c|c|c|} \hline
{\bf Test} & {\bf User time} & {\bf System time} & {\bf BDD nodes} & {\bf Total number of vars} \\ \hline \hline
\multicolumn{5}{|c|}{Full specification}\\ \hline
-reorder & 26487.25s & 420.31s & 34315278 & 47\\ \hline 
\multicolumn{5}{|c|}{Throttle properties}\\ \hline
Base & 5.67s & 0.2s & 112687 & 26\\
-f -inc & 11.06s & 0.07s & 18714 & 26\\
-reorder & 5.69s &  0.15s & 75356 & 26\\
-f -inc -reorder & 12.28s & 0.1s & 18630 & 26\\ \hline
\multicolumn{5}{|c|}{All but throttle}\\ \hline
Base     & 8102.76s  & 50.59s  & 9780674 & 43\\
-f -inc  & 18783.8s  & 263.46s & 3406360 & 43\\
-reorder & 4237.56s  & 7.37s   & 4425504 & 43\\
-f -inc -reorder & 13599.5s & 12.54s & 3406360 & 43\\
\hline
\end{tabular}
\caption{Verification of the Cruise Control specification:  experimental results.
\label{exp_results}}
\end{center}
\horline
\end{table*}}
We were successfully able to model and verify the automobile cruise
control system using the \scrrr\ toolset.  The verification was
performed on a moderately loaded SPARCstation-20 (2 75 MHz processors)
with 256 megabytes of RAM using {\tt SMV} version 2.4.  The complete
Cruise Control specification consists of 22 monitored and 13 controlled
variables, translated into 47 SMV variables.  Given that the
size of the model grows exponentially to the number of variables in
the system, we were not surprised that the initial
runs of {\tt SMV} did not complete after two days.
However, we explored various ways to reduce the time and memory requirements
necessary to verify the system.  The following is the summary of our findings.

\noindent
$\bullet$ The most effective technique is ``slicing'', i.e., removing
variables that are not used in properties under verification.
Currently, if a variable does not appear in any of the properties and
no other variables depend on it, {\tt SMV} still includes it into the
models it builds.  However, it is safe to remove such variables from
consideration, thus greatly reducing sizes of the models.  In SCR,
there is a hierarchy of dependencies between requirements' variables,
which is very easy to compute and take advantage of, although at the
expense of producing separate {\tt SMV} models.  In \scrrr, the later
process can only be done by hand so far, although we are 
developing an automatic model generation.

\noindent
$\bullet$ Another technique is turning the variable reordering on.
{\tt SMV} uses binary decision diagrams (BDDs) to quickly manipulate
logical expressions.  Unfortunately, the size of the BDDs is extremely
sensitive to the order of the variables used to build
them~\cite{atlee96a}.  {\tt SMV} has an option\footnote{Version 2.5 of
{\tt SMV} uses variable reordering by default.}{\tt --reorder} that
allows it to heuristically compute and use a ``better'' variable
reordering which is typically very effective.  However, even with
reordering turned on, {\tt SMV} took a long time to verify the Cruise
Control specification (26487.25 seconds of user time).  To overcome
this problem, we split the model into two: mode classes and controlled
variables related to the throttle, and mode classes and all other
controlled variables.  Results of our experiments appear in
Table~\ref{exp_results}.  In this table, we list the user and the
system time in seconds, and the total number of BDD nodes used during
the verification, as reported by {\tt SMV}.  The last column of the
table is the number of variables in the SMV model.  We were unable to
measure the running time in terms of real time, since {\tt SMV} does
not keep this information.

\noindent
$\bullet$ A more controversial technique is building the state space
incrementally, removing unreachable parts of the model ({\tt SMV}
options {\tt --f} and {\tt ---inc}).  This technique reduces the size
of BDDs at the expense of the longer running time.  Combining this
option with {\tt --reorder} typically yields a smaller number of BDD
nodes than either of the options alone, but the verification time is
slightly worse than by running {\tt SMV} with with {\tt --reorder}
alone (see Table~\ref{exp_results}).  We feel that both options should
be turned on on a fast machine with a limited amount of memory,
whereas just {\tt --reorder} should be turned on on a slower machine.

\begin{table*}[t]
\begin{center}

\begin{tabular}{|l|l|l|l|l|} \hline
\multicolumn{5}{|c|}{\bf Throttle Setting} \\ \hline \hline
{\bf Mode} & \multicolumn{4}{|l|}{\bf Condition} \\ \hline
{\tt Cruise}   & {\tt speed\_slow} & {\tt speed\_ok} \&       & {\tt speed\_fast} \& & {\tt Accel} \& \\
\              & \                 & $\sim${\tt Accel}        & $\sim${\tt Accel}    & $\sim$ {\tt speed\_slow}\\\hline
{\bf Throttle} & {\tt Accel} & {\tt Maintain} & {\tt Decel} & {\tt Off} \\ \hline
\end{tabular}
\caption{Condition table for variable {\tt Throttle}.\label{throttle}}
\end{center}
\horline
\end{table*}
The verification effort
yielded a number of errors, most of which could be traced back to the
original specification~\cite{kirby88}.  We found no errors in the mode
transition table because this table has been analyzed earlier by Atlee
and her colleagues~\cite{atlee96}\footnote{However, our study showed
that a number of conditions used in Atlee's specification were not
necessary, e.g., specifying that transitions from {\tt Override} to
{\tt Cruise} occur only when {\tt Running} and {\tt Ignition} are
true.  These conditions are implied by other transitions in the mode
table and did not need to be specified explicitly.}.  However, we did
find errors in controlled variable tables of the original
specification.  For example, the cruise control system includes a
controlled variable {\tt OilOn} which represents lighting up an
indicator when the vehicle is due for an oil change, i.e., it has
traveled a certain distance since its previous oil change.  In
Kirby's specification, this controlled variable is evaluated when we
enter any mode other than {\tt Off} ({\tt @T(Inmode)} when {\tt
Omiles\_low}).  Thus, the vehicle could be traveling with the cruise
control system turned off, not changing its modes and never
setting {\tt OilOn} to true.  We modified the table for {\tt OilOn} to
set the variable to true when {\tt Omiles\_low} becomes true,
regardless of the mode.

Another problem was found in the table for evaluating the throttle.
In the notation used by Kirby, {\tt Throttle} was an enumerated
variable whose values are evaluated in mode {\tt Cruise} as specified
in Table~\ref{throttle}.  We found that condition {\tt $\sim$Accel} is
redundant -- the system exits mode {\tt Cruise} as soon as {\tt Accel}
becomes true.  We also suspected that the throttle might never become
{\tt Off}.  To check that, we modeled the variable {\tt Throttle} in
{\tt SMV} and ran it against the property $$\sim{\tt Throttle=Off}
\rightarrow AG(\sim{\tt Throttle=Off})$$ which was verified,
confirming our suspicion.  Finally, the correctness of the
specification is conditional upon the time when the predicate on the
vehicle's speed is evaluated.  It works correctly only when {\tt
speed\_ok}, {\tt speed\_slow} and {\tt speed\_fast} get evaluated
{\em before} the system transitions into mode {\tt Cruise}.

\section{Checking Correctness of Code}
\label{codeanalysis}
At some point in the future it will be possible to generate code
from the black-box requirements specified in SCR.  However,
this code will likely require some hand-tuning, e.g., because it is
too slow, or might even have to be  rewritten from scratch.
Mathematically precise software requirements, like SCR,
can be used to reason about correctness of implementations.
\scrrr\ incorporates a tool called {\tt cord}
~\cite{chechik96c} that takes specially annotated source code and
an SCR specification and checks for the correspondence between them.
{\tt cord} uses data-flow analysis instead of exhaustive state
enumeration to enable effective verification in low-degree polynomial
time.  However, the analysis can sometimes be imprecise, i.e.
``there {\em may be} a problem on this line of the code''.
In this section, we describe how to (correctly) annotate
the code, outline the algorithms used in {\tt cord} to check
consistency between requirements and annotated code, and
present results of verifying the implementation of the Cruise
Control System.
\begin{figure*}[t]
\begin{center}
\begin{minipage}[htb]{5in}
\psfig{figure=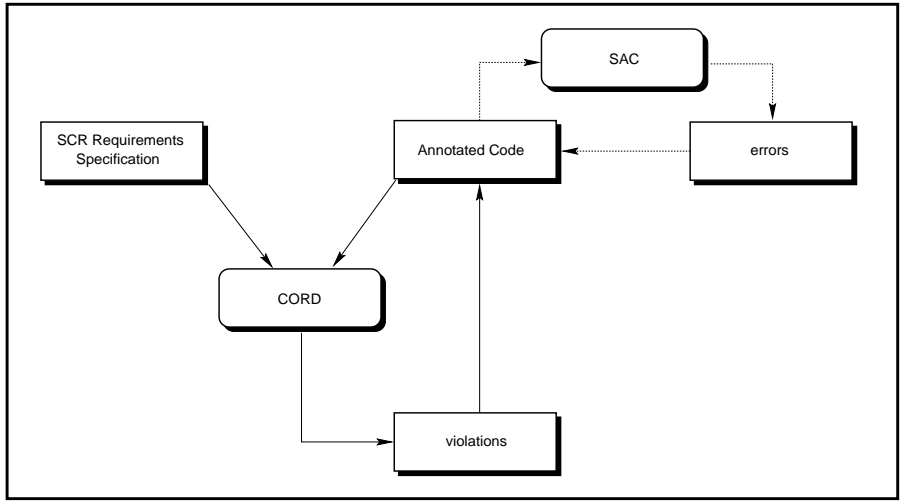,width=4.7in}
\end{minipage}
\caption{Code analysis with \scrrr. \label{fig_cord}}
\end{center}
\horline
\end{figure*}
\subsection{Annotations}
Code is annotated with special statements which describe changes in
the system state.  These changes involve local (rather than invariant)
information and therefore are relatively easy to specify.  The
annotations, described in detail in \cite{chechik96c}, are of three
types.  {\em Initial} annotations indicate the starting states of the
program.  They unconditionally assign values to variables and
correspond to initialization information specified in the
requirements.  {\em Update} annotations assign values to controlled,
monitored or mode class variables\footnote{Monitored and controlled
variables have boolean values; mode class variables have values which
form enumerated types whose constant values are modes of the mode
classes.}.  These annotations identify points at which the program changes
its state.  {\em Assert} annotations assert that  variables
have particular values in the current system state.  This makes our
analysis more precise and serves as documentation of what the developer
assumes about the system at a given point in the program.

Consider the following annotated code fragment, taken from the
implementation of the Cruise Control system:\footnote{Lines that start with
@@ indicate annotations.}
{\small \begin{verbatim}
    if (!vIgnition) {
      @@ Assert ~Ignition;
      vIgnition = true;
      @@ Update Ignition;
      vCC = mInactive;
      @@ Update CC=Inactive;
    }
    else {
      @@ Assert Ignition;
      vIgnition = false;
      @@ Update ~Ignition;
      vCC = mOff;
      @@ Update CC=Off;
    }
\end{verbatim}}
In this fragment, code variables {\tt vIgnition} and {\tt vCC}
correspond to requirements variables {\tt Ignition} and {\tt CC}.
In the Then branch of the If statement we assert that {\tt Ignition}
is false, register that {\tt Ignition} becomes true,
marked by an Update annotation, and then change the mode of the system
to {\tt Inactive}.  This is also marked by an Update annotation.
The Else branch is similar.

As stated above, {\tt cord} uses annotations and control-flow
information of the code to check it for correctness.  That is, ``the
analysis is as good as the annotations''.  Although annotations are
easy to insert, we found ourselves frequently making mistakes, and
also noticed that code maintenance greatly reduces the correspondence
between the code and the annotations.  Thus, the \scrrr\ toolset
includes a tool called {\tt sac}~\cite{chechik98c} designed to check
that the code is annotated correctly.  To use the tool, the programmer
creates a list of {\em correspondences} between variables in the
requirements and the code.  For example, the following is the
specification of correspondences for the above code fragment: 
{\small
\begin{verbatim} 
   correspondences: 
   {Ignition} -> {vIgnition}; 
   {CC} -> {vCC};
\end{verbatim}}
These correspondences can be one-to-one (one requirements variable
corresponds to one code variable), one-to-many (one requirements
variable corresponds to several code variables), and many-to-one.
{\tt sac} ensures that an assignment to (a check of) a code variable
is always followed by an Update (an Assert) of an appropriate
requirements variable.  The entire process of code verification with
\scrrr\ is depicted in Figure~\ref{fig_cord}. Here, tools ({\tt sac}
and {\tt cord}) and artifacts are represented by ellipses and boxes,
respectively.

\subsection{Analysis}

Analysis done by {\tt cord} is described in detail
elsewhere~\cite{chechik96c}.  In this section, we give a quick summary
of this process.

{\tt cord} checks that transitions implemented in the code are exactly
the same as those specified in the requirements.  These checks
correspond to three types of properties: (1) the code and the
specification start out in the same initial state; (2) the code
implements {\em all} specified transitions; and
(3) the code implements {\em only} specified transitions.
Properties of types (2) and (3) are called ALT (``all
legal transitions'') and OLT (``only legal transitions''),
respectively.  For example,  {\tt cord} checks an ALT property that a transition from mode {\tt Cruise}
to mode {\tt Off} on event {\tt @F(Ignition)} exists in the code.
This transition refers to the fourth row of Table~\ref{table_CC}.  One of the OLT
properties is ``the only transitions into mode {\tt Off} are from mode
{\tt Inactive} on event {\tt @F(Ignition)}, or from mode {\tt Cruise}
on event {\tt @F(Ignition)}, or from mode {\tt Override} on event {\tt
@F(Ignition)}''.

Verification is done via static analysis of the annotated code.
A technique similar to {\em constant propagation} is used to
create a finite-state abstraction of the annotations and control-flow
program statements of the program.  We use an aggressive state-folding
strategy aimed at minimizing the number of states.  This number is
bound above by the number of annotations and control-flow structures
in the code and is almost not affected by the number of variables
in the specification.  After the model has been created, we check 
it for consistency with the specification.  Typically, the properties
involve fairly short (2-3 states) fragments of the paths through
the code, thus enabling very efficient checking.  In addition, {\tt cord}
can verify invariant and reachability properties, find unreachable
states, and check that environmental assumptions are not violated in the
code.  

Fast and highly scalable processing comes at a price of inexact verification.
Abstraction used in {\tt cord} leads to computing more
behaviors than can be present in the code.  Thus, OLT properties and
invariants are checked {\em pessimistically}, i.e., some violations that
are not present in the code can be reported.  ALT and reachability properties
are checked {\em optimistically}, i.e., some violations in the code can be
overlooked.  We believe that {\tt cord} should be used as a debugging rather
than a verification tool and found it to be extremely effective in discovering
errors (see Section~\ref{checkcc}).

\begin{figure}[htb]
\begin{center}
\begin{minipage}[htb]{3in}
\psfig{figure=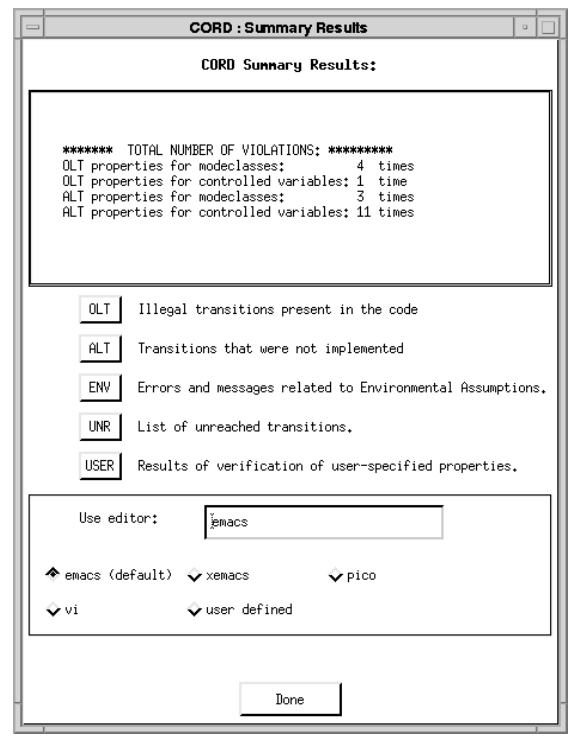,width=2.9in}
\end{minipage}
\caption{Summary of errors.  \label{fig-errors}}
\end{center}
\end{figure}
\begin{figure}[htb]
\begin{center}
\begin{minipage}[htb]{3in}
\psfig{figure=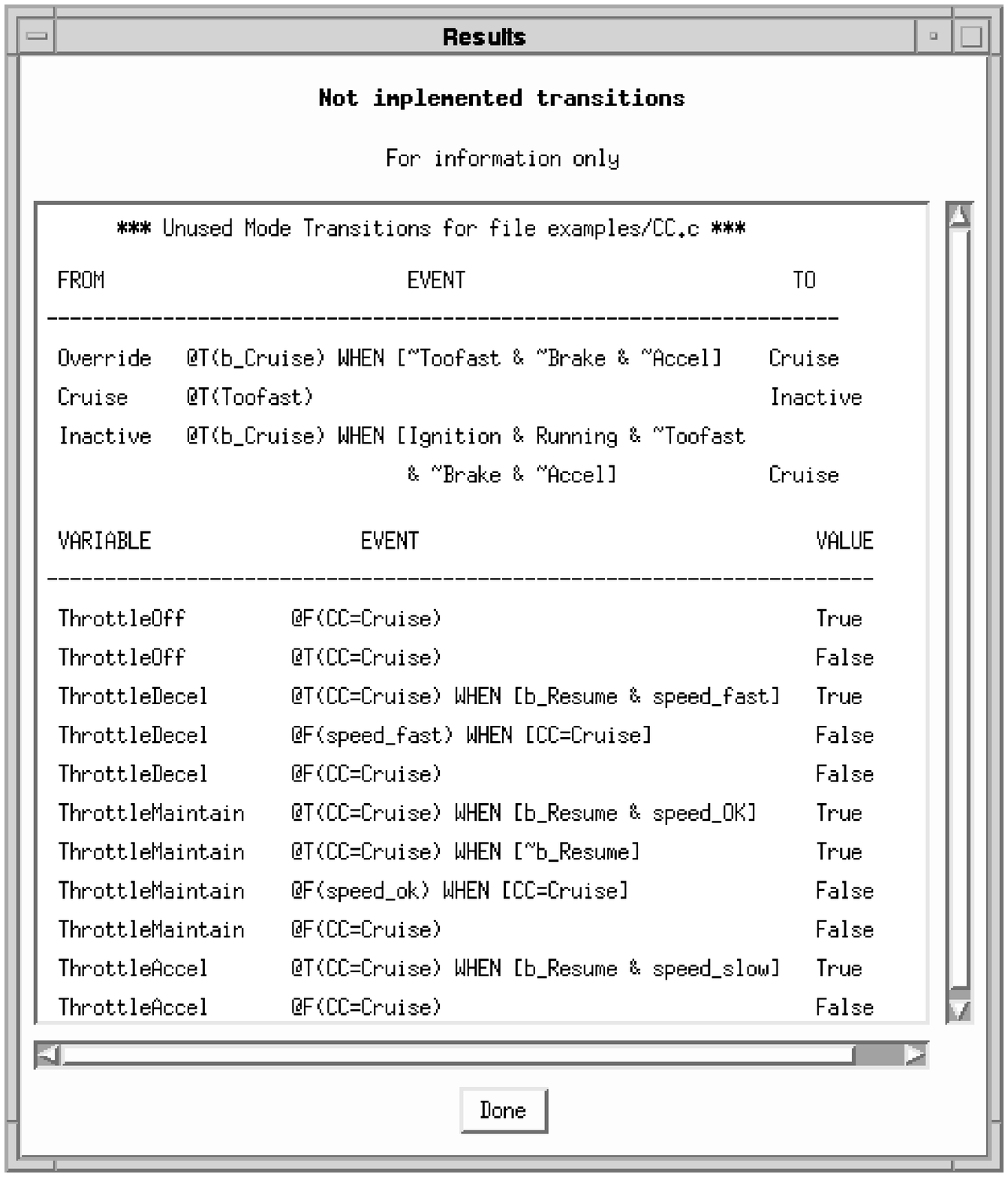,width=2.7in}
\end{minipage}
\caption{ALT errors.  \label{fig-alt}}
\end{center}
\end{figure}

We had to resort to annotating code because our analysis techniques
could only handle simple operations on booleans and enumerated types.
A next generation of {\tt cord} is currently being developed.
For finite types, this version of the tool will be able to handle
more interesting operations, like addition and comparison.  It will
also be able to approximate infinite types and perform operations
on them using abstract interpretation~\cite{cousot77}.  This will allow
us to analyze code directly rather than using annotations.

\subsection{Checking the Cruise Control System}
\label{checkcc}
We have implemented the Cruise Control system in C with
an Xlib interface.  The implementation was not originally annotated
and consisted of 675 lines of code, out of which  roughly 380 lines were used
to implement the GUI of the system.    The code was then annotated by
another member of the group in about 40 minutes; the annotation effort was
trivial and resulted in 37 Update, 25 Assert and 1 Initial annotation.

\begin{figure*}[t]
\begin{center}
\begin{minipage}[htb]{4.2in}
\psfig{figure=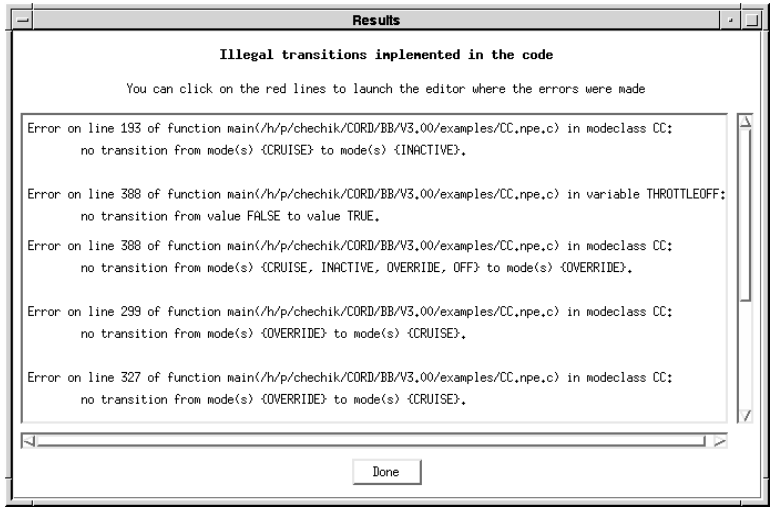,width=4in}
\end{minipage}
\caption{OLT errors.  \label{fig-olt}}
\end{center}
\horline
\end{figure*}
After fixing annotation errors reported by {\tt sac}, we ran {\tt
cord} on the code of the Cruise Control system.  The analysis took 7.62
seconds (3.45 user, 0.74 system, as reported by tshell's {\tt time}
command) on a moderately loaded SPARCstation-4 (85 MHz processor) with
64 megabytes of RAM, and resulted in reporting 14 ALT and 5 OLT violations (see Figure~\ref{fig-errors}).
The ALT and the OLT violations are shown in Figures~\ref{fig-alt} and \ref{fig-olt}, respectively.

One OLT and one ALT violation were caused by an incorrect annotation
overlooked by {\tt sac}.  {\tt sac} does not consider values of
variables when checking code for correct annotations, and did not
report a violation when we annotated an assignment with {\tt Accel}
instead of {\tt $\sim$Accel}.  Another pair of OLT and ALT violations
came from an incorrect transition in the code (line 388) -- to mode {\tt
Override} instead of {\tt Cruise}.  In addition, {\tt cord} computed
that the system can be in all possible modes before this line, whereas
only {\tt Cruise} is possible.  This problem arises because of
the imprecise analysis used in {\tt cord} and can be fixed by adding an
extra Assert annotation.  A yet another pair of OLT and ALT violations
came from a transition from {\tt Override} to {\tt Cruise}.  The code
did not check that this transition is only enabled when {\tt Toofast},
{\tt Brake} and {\tt Accel} are false.  Out of the remaining two OLT
violations, one was a false negative and the second was an incorrect
triggering event for the variable {\tt ThrottleOff}.  All other ALT
properties came from transitions that were not implemented in the
code.


\section{Conclusion}
\label{conclusion}

In this paper we presented \scrrr\ -- an integrated toolset for
specification and reasoning about tabular requirements.  Through a
unified interface, \scrrr\ allows to check software requirements for
correctness and analyze consistency between annotated code and
requirements.  We are currently working on developing support for automatic
code generation and are looking into ways of using SCR for generation of
black-box test cases.
We believe that \scrrr\ is an important step towards
increasing usability of formal methods: it attempts to replace
free-form reasoning in logic by easy to write and review structured
tables and amortizes the cost of creation of formal requirements
through multiple automated analysis activities.  We envision the
following methodology for using \scrrr: 
\begin{enumerate}
\item A (human) requirements designer provides an SCR tabular specification
of the system's required behavior and a set of global properties that should be
satisfied by the system.

\item \scrrr\ automatically translates the SCR specification into the input
to a model-checker, verifies properties and translates counter-examples
into the SCR notation.
\item Once the specification is considered correct, \scrrr\ generates
an implementation of the mode logic and stubs for interface code, which
are filled in by a human developer.  
\item Using the specification, \scrrr\ generates test cases.  Although
the mode logic is correct by construction, interface is likely to need
testing.
\item The resulting implementation is analyzed for performance
and optionally manually fine-tuned.
\item If manual fine-tuning was used, the code is automatically checked 
for consistency with its specifications.
\item Once the problems pointed out by the analysis are fixed, the code
is tested using (a subset of) test cases generated in step (4) above.
\end{enumerate}
\begin{figure*}[t]
\begin{center}
\begin{minipage}[htb]{5in}
\psfig{figure=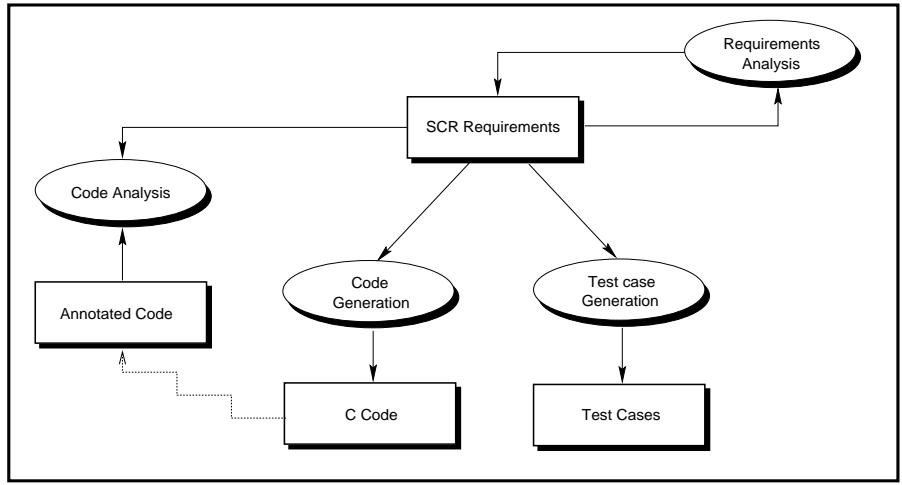,width=4.7in}
\end{minipage}
\caption{Pictorial description of activities that can be performed with \scrrr. \label{bigpicture}}
\end{center}
\horline
\end{figure*}
In Figure~\ref{bigpicture}, analysis activities
performed within \scrrr\ and software artifacts are shown inside ellipses
and boxes, respectively.  For example, the code generation activity
takes SCR requirements as an input and generates an implementation in
C.  A dashed line signifies that the generated code and the code that
is analyzed for correctness, represent the same artifact.

In \scrrr, requirements analysis is done via model-checking.  Although
model-checking is an effective verification technique which can be
used without any user input, it has a number of limitations.  Checking
can often be prohibitively expensive and cannot usually be
applied to infinite-state systems.  Thus, we are limited to verifying
just control (as opposed to data or timing) aspects of the
specification.  In order to use model-checking effectively, we had to
reduce the expressive power of our input logic, which may not be
feasible for many applications.  We recognize that more sophisticated
verification might be necessary and are planning to experiment with
using a theorem-prover, e.g.  PVS~\cite{owre93}, to check complex typechecking
and timing properties of systems.

Code verification in \scrrr\ is done via a static-analysis tool {\tt cord}.
Although effective in finding errors in our case studies, {\tt cord}
is still a prototype tool in need of major improvements.   We are
currently redesigning it to process a more expressive annotation
language and reason about infinite-domain variables.  This would
make {\tt cord} able to verify implementations of unrestricted
SCR specifications.


\vspace{4mm}

\noindent
{\Large\bf Acknowledgments}

\vspace{4mm}

\noindent
Joanne Atlee and her students developed {\tt gensmv} and performed the
initial analysis of the Cruise Control system.  They also helped in
shaping the look and feel of the GUI for specifying requirements.  The
author is also grateful to John Gannon and Rich Gerber for teaching her about
formal methods, to Connie Heitmeyer and Stuart Faulk for many fruitful
discussions about SCR, to David Wendland and Hicham Mouline who
implemented and improved many aspects of the tools, to Sai Sudha for
developing {\tt sac}, and to Matthew Cwirko-Godycki for his help with the
Cruise Control case study.

This research was supported in part by the NSERC Grant \# OGP0194371 and
by the University of Toronto's Connaught Fund.

\vspace{4mm}

\noindent
{\Large\bf About the Author}

\vspace{4mm}

\noindent
Marsha Chechik is an assistant professor in the Department of Computer
Science at the University of Toronto.  She can be reached at the
address Department of Computer Science, University of Toronto, 10 King's
College Rd., Toronto, ON M5S 3G4.  Her e-mail address is {\tt chechik@cs.toronto.edu}.

Dr. Chechik's interests are mainly in the application of formal methods
to improve quality of software.  She is also interested in other aspects
of software engineering and in computer security.

\end{document}